\input{aipcheck}

\documentclass[
    ,final            % use final for the camera ready runs
  ]
  {aipproc}

\layoutstyle{6x9}

\usepackage{amssymb}
\usepackage{graphicx,natbib}

\newcommand{\be}{\begin{equation}}
\newcommand{\ba}{\begin{eqnarray}}
\newcommand{\ee}{\end{equation}}
\newcommand{\ea}{\end{eqnarray}}

\begin{document}

\title{The Theory and Simulation of the 21-cm Background from
the Epoch of Reionization}



\author{Paul R. Shapiro}{
  address={Department of Astronomy, University of Texas, Austin, TX 78712-1083,
  U.S.A.}
}

\author{Ilian T. Iliev}
{ address={Institute of Theoretical Physics, University Zurich,
Winterthurerstrasse 190,
CH-8057 Zurich, Switzerland}
}

\author{Garrelt Mellema}{
  address={Stockholm Observatory, AlbaNova
  University Center, Stockholm University, SE-106 91 Stockholm, Sweden}
}

\author{Ue-Li Pen}{
  address={Canadian Institute for Theoretical Astrophysics, University
  of Toronto, 60 St. George Street, Toronto, ON M5S 3H8, Canada}
}
\author{Hugh Merz}{
  address={Department of Physics \& Astronomy, University of Waterloo, 200 University Avenue West, Waterloo, ON N2L 3G1, Canada} }

\begin{abstract}
	The redshifted 21-cm line of distant neutral H atoms provides 
a probe of the cosmic ``dark ages'' and the epoch of reionization (``EOR'')
which ended them, within the first billion years of cosmic time.
The radio continuum produced by this redshifted line can be seen in 
absorption or emission against the cosmic microwave background (``CMB'')
at meterwaves, yielding information about the thermal and ionization 
history of the universe and the primordial density perturbation spectrum
that led to galaxy and large-scale structure formation. Observing this 
21-cm background is a great challenge, as it is necessary to detect 
a diffuse signal at a brightness temperature that differs from that of the CMB at 
millikelvin levels and distinguish this from foreground continuum sources.   
A new generation of low-frequency radio arrays is currently under 
development to search for this background.  Accurate theoretical predictions
of the spectrum and anisotropy of this background, necessary to guide and
interpret future observations, are also quite challenging.  Toward this
end, it is  necessary to model the inhomogeneous reionization of the
intergalactic medium and determine the spin temperature of the 21-cm 
transition and its variations in time and space as it decouples from the 
temperature of the CMB.  In my talk, I summarized some of the theoretical
progress in this area.  Here, I will focus on just a few of the predictions
for the 21-cm background from the EOR, based on our newest, large-scale
simulations of patchy reionization.  These simulations are the first with
enough N-body particles (from 5 to 29 billion) and radiative transfer rays 
to resolve the formation of and trace the ionizing radiation 
from each of the millions of dwarf galaxies believed responsible for
reionization, down to $10^8 M_\odot$, in a cubic volume 
large enough ($90$ and $163$ comoving Mpc on a side) to make meaningful
statistical predictions of the fluctuating 21-cm background.    
\end{abstract}

\maketitle

\section{Introduction} 

The redshifted 21-cm line of neutral hydrogen atoms 
can probe the thermal and ionization history of the universe
from the cosmic ``dark ages'' through the EOR and the density
perturbations responsible for structure formation 
(see, e.g., \cite{2006PhR...433..181F} for review and refs). 
   It is generally believed that, during the EOR,
the spin temperature, $T_s$, of this hyperfine transition was decoupled
from the CMB temperature, $T_\mathrm{CMB}$, by the Wouthuysen-Field
mechanism, whereby UV photons at the Ly $\alpha$ transition frequency are
resonantly scattered, and the excited atoms then decay to a different
hyperfine level of the ground state.  This radiative pumping
drives $T_s$ toward the gas kinetic temperature, $T_K$.  The 21-cm line
is seen in emission or absorption against the CMB if $T_s$ is greater
than or less than $T_\mathrm{CMB}$, respectively.  If, as is also generally
believed, other radiation is also present (e.g. X-rays) to heat the neutral
IGM so that $T_K \gg T_\mathrm{CMB}$, then the 21-cm differential
brightness temperature depends only upon the
neutral gas density and peculiar velocity fields, independent of the value of $T_K$ or $T_s$.
Henceforth, we shall limit our discussion of the 21-cm background from
the EOR to this regime.

Reionization is generally believed to be the outcome of the release of
ionizing radiation by galaxies undergoing star formation (see,
e.g., \citep{2006Sci...313..931B,2005SSRv..116..625C} for recent
reviews). Current theory suggests that the galaxies responsible for
most of this radiation are dwarf galaxies more massive than about
$10^8M_\odot$. 
While dark matter dominates the gravitational
forces which cause this structure formation, ordinary atomic matter
must join the dark matter in making galaxies for star formation to be
possible.  Once the atomic gas in the intergalactic medium (``IGM'') in
some region is heated by reionization, however, gas pressure opposes
gravitational collapse, and, thereafter, the smallest galaxies form
without atomic matter and cannot make stars. The minimum mass of
star-forming galaxies in such regions is about $10^9$ solar
masses. This suppression of star-forming galaxies of lower mass inside
ionized regions is sometimes referred to as ``Jeans-mass filtering''
\citep{1994ApJ...427...25S}.

Due to its complexity, reionization is best
studied through large-scale numerical simulations
\citep{2006MNRAS.369.1625I, 2007ApJ...654...12Z, 2007ApJ...657...15K,
  2007ApJ...671....1T}, combining the numerical challenges of $N$-body
simulation and radiative transfer.  The tiny galaxies which are the
dominant contributors of ionizing radiation must be resolved in very
large cosmological volumes, large enough to contain billions of times
more total mass than one dwarf galaxy and up to tens of millions of
such galaxies, in order to derive their numbers and clustering
properties correctly.
 The expansion of the intergalactic ionization fronts created by all
these millions of galaxies must, in turn,
be tracked with a 3D radiative transfer method which also solves 
non-equilibrium ionization rate equations.
The combination of all these
requirements makes this problem formidable.

Our original simulations \citep{2006MNRAS.369.1625I} resolved the
formation of all galaxies more massive than about one billion solar
masses, those which are expected to form stars even after reionization
has heated their environment. These simulations of a comoving,
cosmologically-expanding cubic volume $(150 \textrm{ Mpc})^3$ enabled
us to make the first statistically meaningful
predictions of the impact of reionization on observables like the
$21 \textrm{-cm}$ background \citep{2006MNRAS.372..679M},
the CMB anisotropy \citep{kSZ}, and the
observability of early galaxies as Lyman $\alpha$ line-emission
sources \citep{2007arXiv0711.2944I}, essential predictions to guide
and interpret current and future observational programs
\citep{2008MNRAS.384..863I}. With the next generation of simulations
described here, we shall explore the role of less massive dwarf
galaxies. These are important sources of ionizing radiation if they
form before their neighborhood is reionized but are prevented from
being sources if they form after it is reionized. Our previous
results showed that the inclusion of these sources and of their
suppression changes the outcome of reionization substantially 
\citep{2007MNRAS.376..534I}.    
Simulations of volume $(50 \textrm{ Mpc})^{3}$ showed that
reionization was ``self-regulated,'' i.e. the more abundant low-mass
halos initially dominated but later choked themselves off as their
formation was suppressed in ionized regions.  The next step, described
here, is to push these simulations of ``self-regulated'' reionization
to larger volumes, large-enough to make meaningful statistical
predictions of observables like the 21-cm background.

\section{New, Large-scale Simulations of Self-Regulated Reionization}

We have developed a new, cosmological $N$-body code more
powerful than the one we used for our previous simulations \cite{teragrid08}. 
CubeP$^3$M,
a massively-parallel P$^3$M (particle-particle-particle-mesh)
\citep{Hockney:1988:CSU} code, is a successor to the previous code, PMFAST
\citep{2005NewA...10..393M}. Its force resolution is significantly better than that of PMFAST, by inclusion of the short-range correction from direct particle-particle (PP) forces. 
With this improved force resolution, dark mater halos can be identified with
as few as 20 particles per halo, to locate all the galaxies which are possible
sources of ionizing starlight, with masses above this minimum. 

These $N$-body results provide the evolving density field of the IGM
and the location and mass of all the halo sources, as input to a
separate radiative transfer simulation of inhomogeneous
reionization. The latter simulation is performed by our C$^2$Ray
(\textbf{C}onservative, \textbf{C}ausal \textbf{Ray}-Tracing) code, a grid-based, ray-tracing,
radiative transfer and nonequilibrium chemistry code, described in
\citep{methodpaper}. The ionizing radiation is ray-traced from every
source to every grid cell using a method of short characteristics. The
code is explicitly photon-conserving in both space and time, which
ensures an accurate tracking of ionization fronts, independent of the
spatial and time resolution, even for grid cells which are optically
thick to ionizing photons and time steps long compared to the
ionization time of the atoms, with corresponding great gains in
efficiency.  The code has been tested against analytical solutions
\citep{methodpaper} and, in direct comparison with other radiative
transfer methods, on a standardized set of benchmark problems
\citep{comparison1}.  C$^2$Ray has also been used in the first, 3D
radiation-hydrodynamical simulations of an H\,{\sc II} region in a turbulent
molecular cloud \citep{2006ApJ...647..397M}. 
While our basic methodology remains essentially as described in
\citep{methodpaper}, our C$^2$Ray code is now 
parallelized for distributed-memory machines \cite{teragrid08}.

With CubeP$^3$M, we simulated the $\Lambda$CDM universe with $1728^3$
and $3072^3$ $N$-body particles of mass $5 \times 10^{6}\ M_{\odot}$,
in comoving box sizes of 91~Mpc and 163~Mpc, respectively.
 At 29 billion particles, the latter is the largest N-body
simulation of the formation of early cosmological structure to date.
This mass resolution allows us to resolve all halos above
$10^8M_\odot$, roughly the minimum mass of halos which can radiatively cool
by hydrogen atomic-line excitation and efficiently form stars.
These halos are separated into halos 
above the Jeans mass of the ionized IGM ($\sim
10^9M_\odot$), which are not affected by reionization, and lower-mass
halos, whose star formation is suppressed if the halo formed inside an
intergalactic H\,{\sc II} region.

\begin{figure}[tbp]
  \includegraphics[height=2.5in]{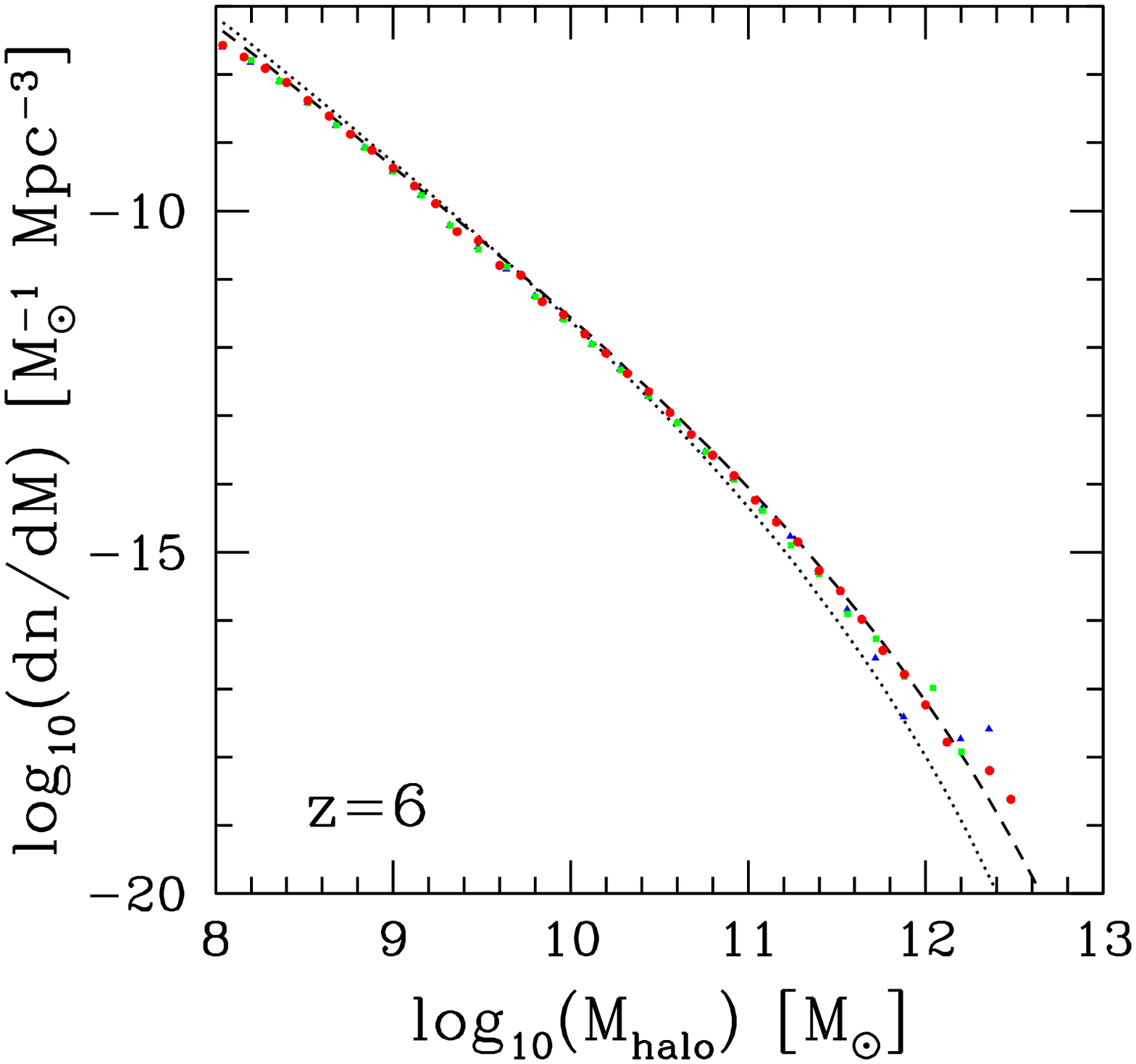}
  \includegraphics[height=2.5in]{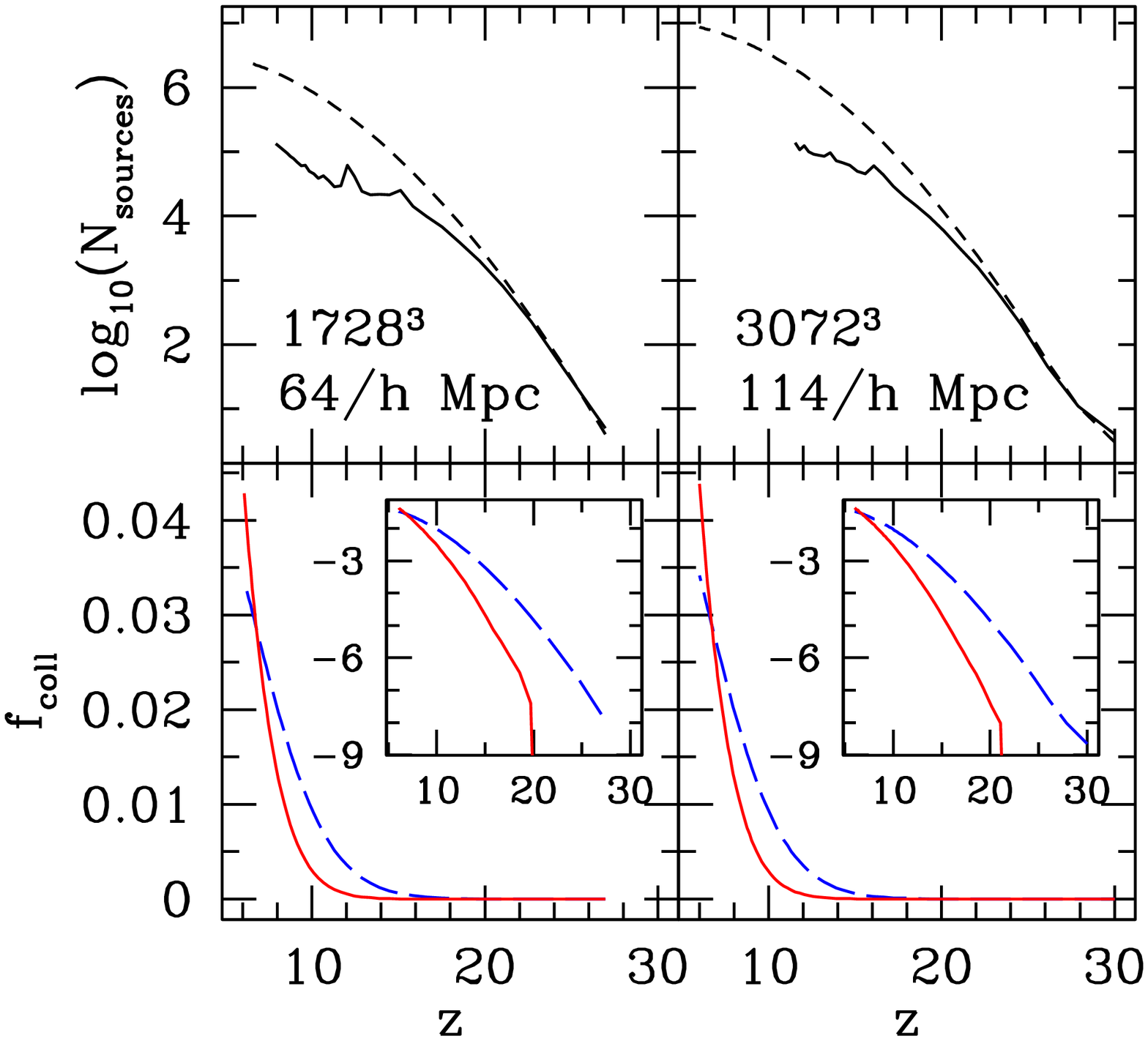}
  \caption{(a)(left) Halo mass functions at $z=6$ from $N$-body simulations 
  of $\Lambda\rm CDM$ with
    box sizes 91~Mpc (blue triangles), 103~Mpc (green
    squares) and 163~Mpc (red circles). The three mass
    functions agree with each other, except for the largest
    halos, which are subject to cosmic variance for the smaller volume
    simulations. The lines show the predicted mass functions based on
    the well-known analytical models of Press-Schechter
    (\citep{1974ApJ...187..425P}; dotted) and Sheth-Tormen
    (\citep{2002MNRAS.329...61S}; dashed). (b) (right)
    (top panels) Total number of halos (dashed) and
    number of active sources (i.e. total number minus the suppressed
    ones; solid); and (bottom panels) collapsed mass fraction in
    high-mass (red, solid) and low-mass (i.e. suppressible) sources (blue, dashed)
    (insets show the same in log scale) vs. cosmic redshift for
    simulations f100\_250S\_91Mpc (left) and f50\_250S\_163Mpc (right).
    \label{massf}}
\end{figure}

Figure~\ref{massf} (a) shows the halo
mass functions at $z=6$ for three of our N-body simulations. 
These $N$-body simulations, i.e.\ the halo catalogues and density
field, are the basis for our C$^2$Ray
radiative transfer simulations of reionization.  All halos
are assumed to host galaxies, and thus, ionizing sources, unless they
form inside an H\,{\sc II} region with a mass below the threshold for
``Jeans-mass filtering''. To each galaxy, we assign an ionizing photon
luminosity proportional to the halo mass, with an efficiency,
$f_\gamma$, which gives the number of photons emitted per atom over
the lifetime of the sources (here assumed to be 17.6 and 11.5 Myr for
our 91 Mpc and 163 Mpc simulations, respectively). The simulations are
denoted f100\_250S\_91Mpc\_432 and
f50\_250S\_163Mpc\_384, where the first two numbers are the values of
$f_\gamma$ for the high- and low-mass sources, respectively, ``S''
indicates that source suppression due to Jeans-mass filtering was
included, the third is the box size and the last is the radiative
transfer grid resolution per dimension.

\begin{figure*}[tbp]
  \includegraphics[width=2.85in]{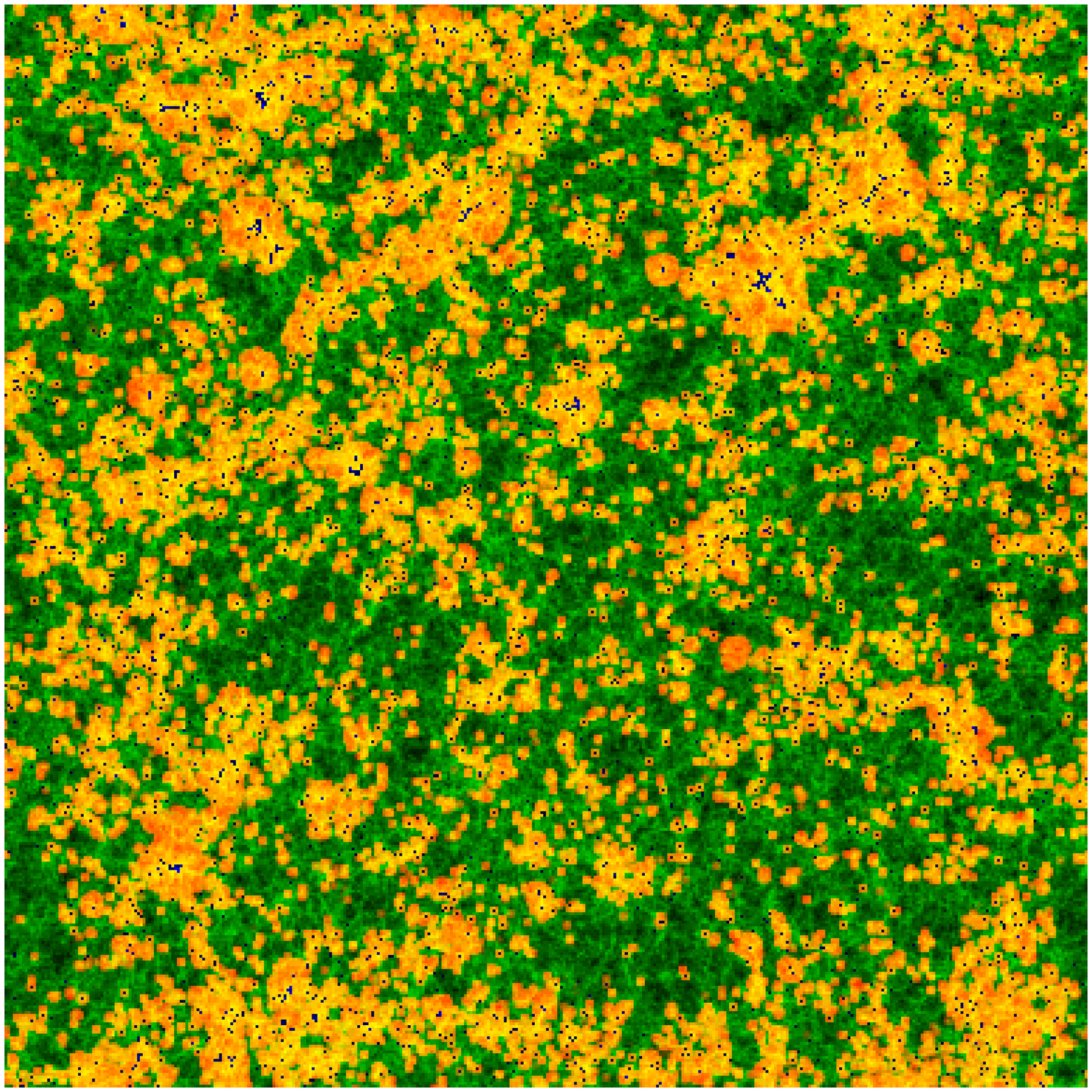}
  \includegraphics[width=2.85in]{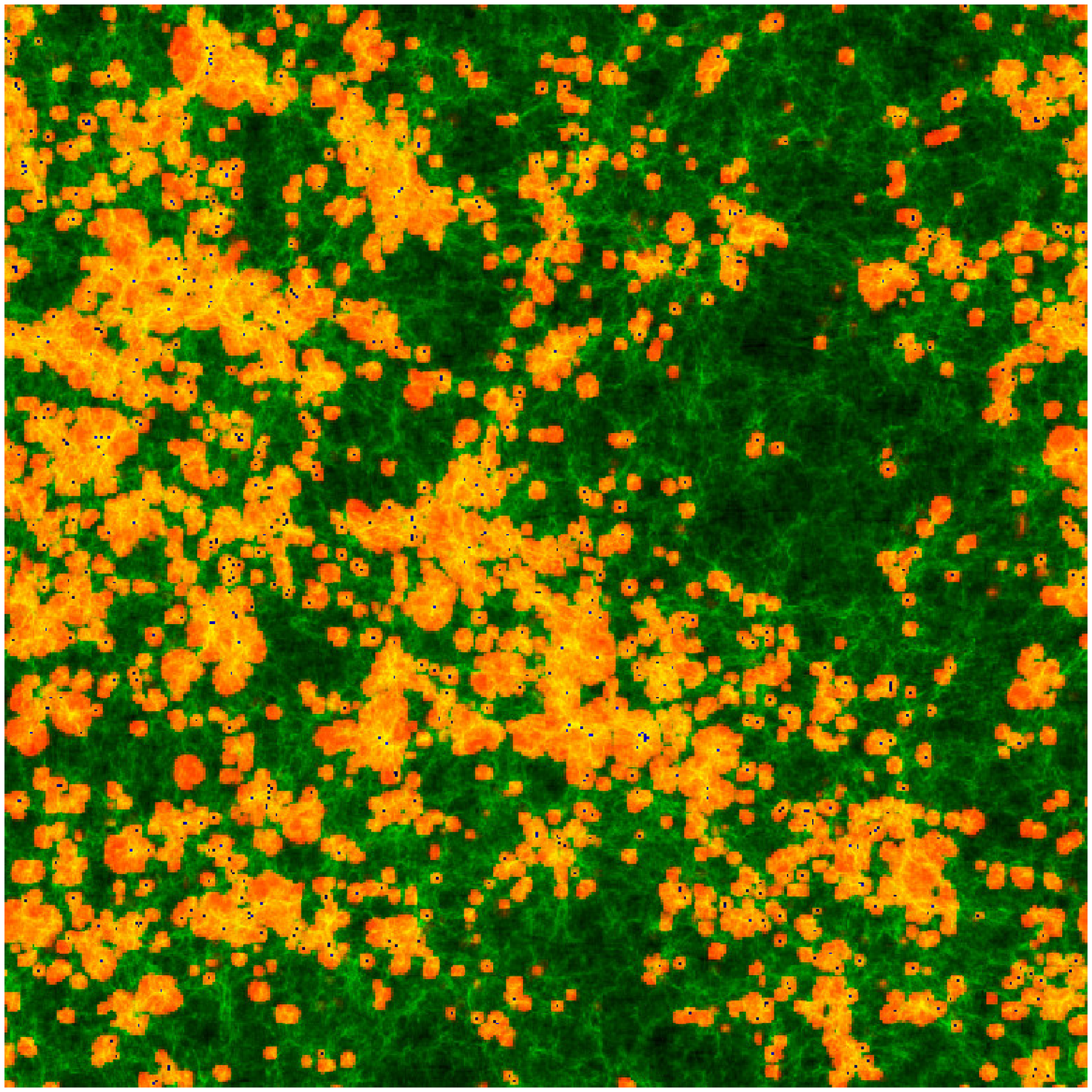} 
  \caption{ Spatial slices
  of the ionized and neutral gas density from our radiative transfer
  simulations (a)(left) boxsize 163~Mpc at redshift $z=11.6$ and
  (b)(right) boxsize 91~Mpc at $z=11.9$, both at box-averaged ionized
  fraction by mass $x_m=0.30$. Shown are the density field (green)
  overlayed with the ionized fraction (red/orange/yellow) and the
  cells containing sources (dark/blue). There is a good agreement
  between the two simulations in the typical sizes of the
  locally-percolated ionization bubbles. [A ``Quicktime'' animation of
  the simulation on the right is included here.]}
\label{images}
\end{figure*}

Figure~\ref{massf}(b) shows the evolution of the total source numbers,
active source numbers and the fractions of the mean mass density
collapsed into the two types of sources, those too massive to be
suppressed by reionization and those which are sources only if they
form in a neutral patch of the IGM.  Figure~\ref{images} (and
accompanying animation\footnote{This Quicktime movie is playable with MPlayer and some Windows Media Players.}) shows images of the gas density field and
(active) source halos overlayed with the ionized fraction field
derived from our radiative transfer simulations. These indicate the
complex, evolving geometry and topology of the network of ionized
bubbles and neutral patches, which we must calculate in detail in
order to make predictions of the fluctuating 21-cm background from
this EOR.  

\begin{figure*}
  \includegraphics[width=2.5in]{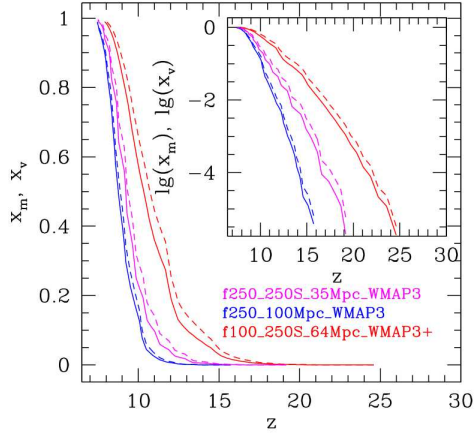}
  \caption{Mean
  ionized fractions, $x_m$ (mass-weighted; dashed) and $x_v$
  (volume-weighted; solid) vs.  redshift for three simulations: new
  simulations of box size 91 Mpc (f100\_250S\_64Mpc\_WMAP 3+; red;
  N.B. size label ``64'' here means ``$64/h$'', where $h=0.7$); previous
  ``self-regulated reionization'' simulations of box size 50 Mpc
  (f250\_250S\_35Mpc\_WMAP3; magenta) from \cite{2007MNRAS.376..534I}; previous
  large-volume simulations \textit{without} low-mass halos, of box
  size 143 Mpc (f250\_100Mpc\_WMAP3; blue) from \cite{2008MNRAS.384..863I}.} \label{mwvw}
\end{figure*}

The evolution of the mean ionized fractions of the IGM, mass-weighted
($x_m$) and volume weighted ($x_v$), is shown in Figure~\ref{mwvw},
including a comparison with those from our previous simulations of
``self-regulated'' reionization in a smaller box ($50$ Mpc), described in
\cite{2007MNRAS.376..534I}, and those from our previous large-box ($143$ Mpc) simulations with
halo masses above $10^{9} M_{\odot}$ only (i.e. no suppressible,
low-mass halos) \cite{2008MNRAS.384..863I}. In all cases, reionization
proceeds ``inside-out'' -- i.e. halos form in high-density regions,
clustered around density peaks and ionize these environments first, so
$x_m > x_v$.  For the new, self-regulated case shown there, $x_m =
0.1, 0.5,$ and $0.99$ at $z = 13.9, 10.7$, and $7.9$, respectively.
As described in \cite{2007MNRAS.376..534I}, the effect of including
the lower-mass (suppressible) halos is to enable reionization to start
earlier, since the low-mass halos are more abundant earlier.  The end
of reionization is not significantly affected, however, since it is
dominated by the exponential rise of the abundance of the higher-mass
halos, which are unaffected by Jeans-mass filtering of the IGM, while
the abundance of low-mass halo sources saturates as the ionized volume
grows.  

\begin{figure*}
  \begin{tabular}{c}
    \includegraphics[width=4.0in, clip]{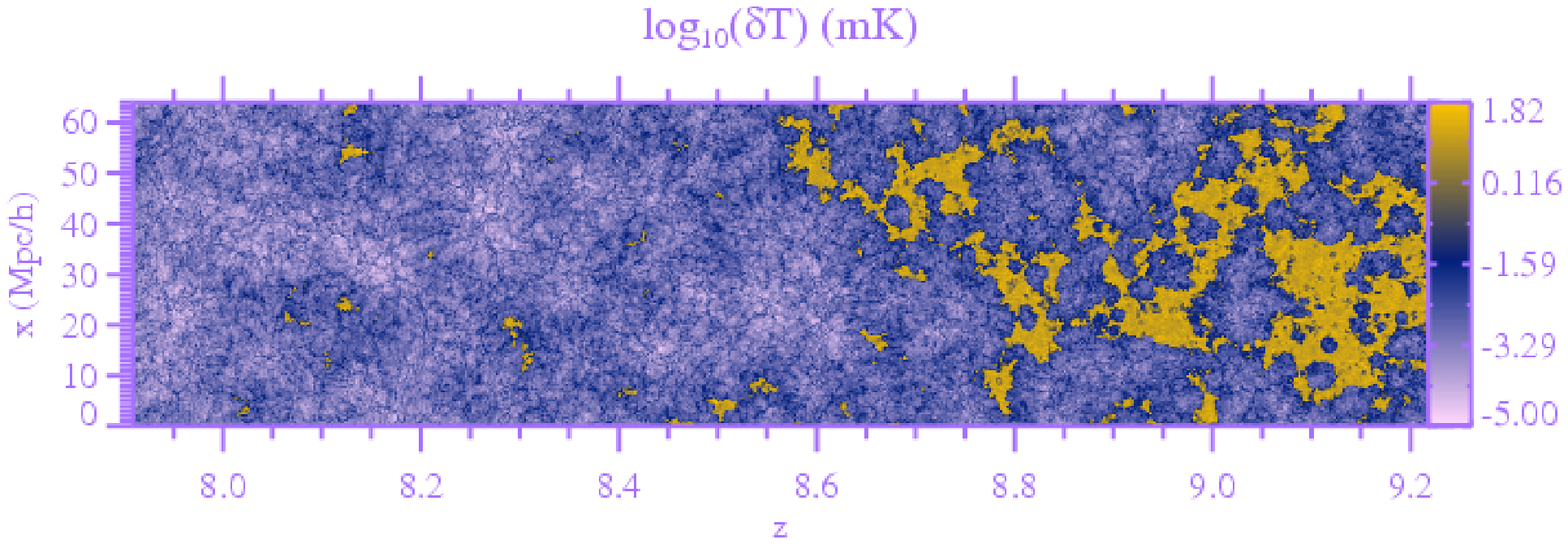} \\
    \includegraphics[width=4.0in, clip]{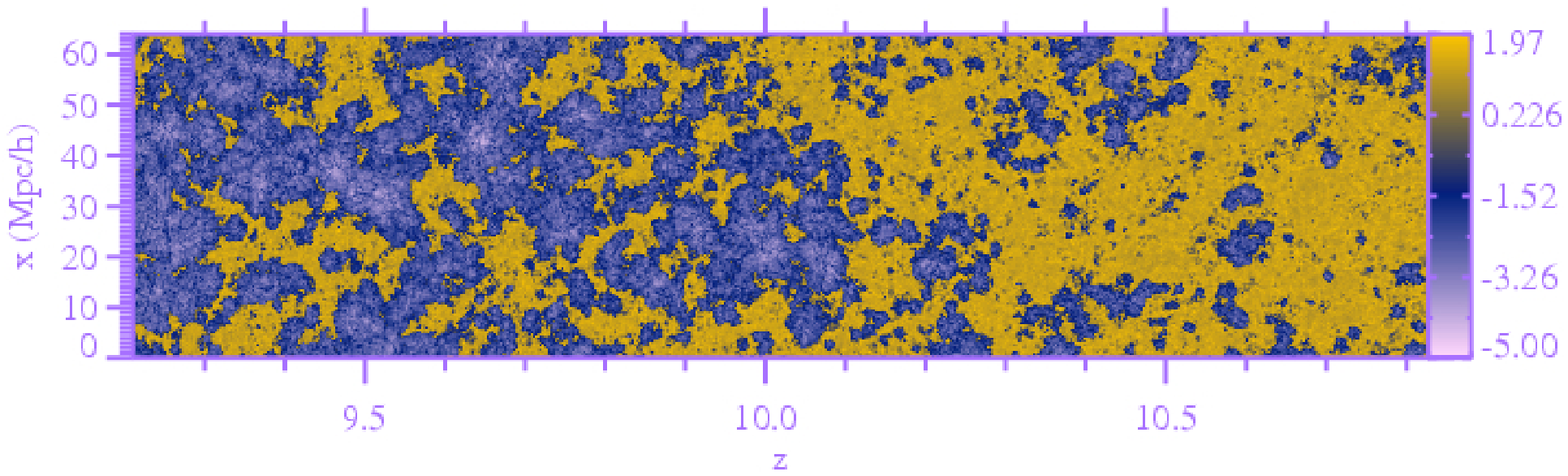}
  \end{tabular}
  \caption{Sky map of $21$-cm brightness temperature fluctuations,
  $\delta T_b \equiv (T_b-T_\mathrm{CMB})/T_\mathrm{CMB}$, along LOS,
  for $91$ Mpc box reionization simulations, at observed frequencies
  $\nu_{obs} (\mathrm{MHz}) = 142 [(1+z)/10]$.}  \label{skymap}
\end{figure*}

\begin{figure}
  \includegraphics[height=1.89in]{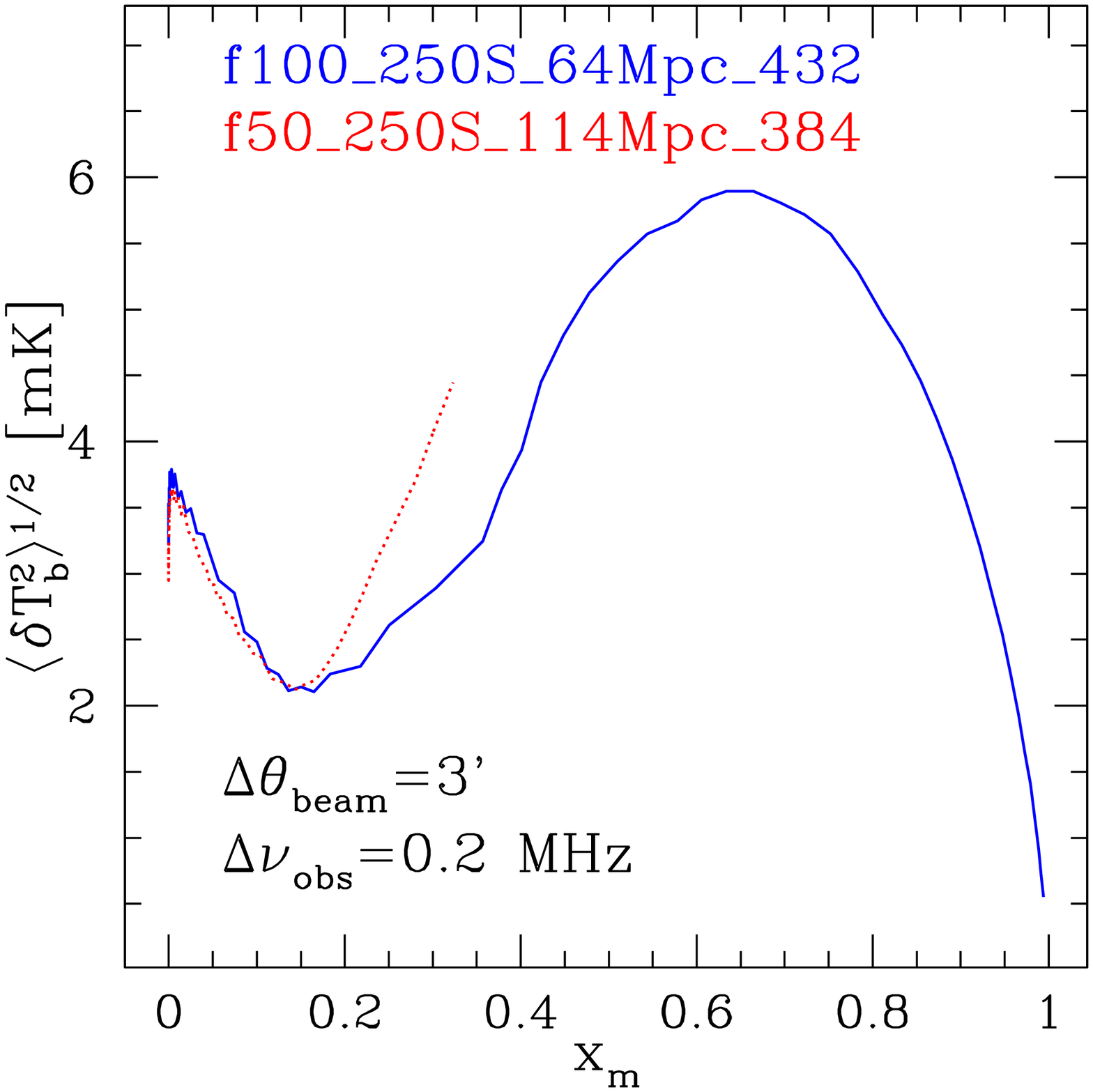}
  \includegraphics[height=1.89in]{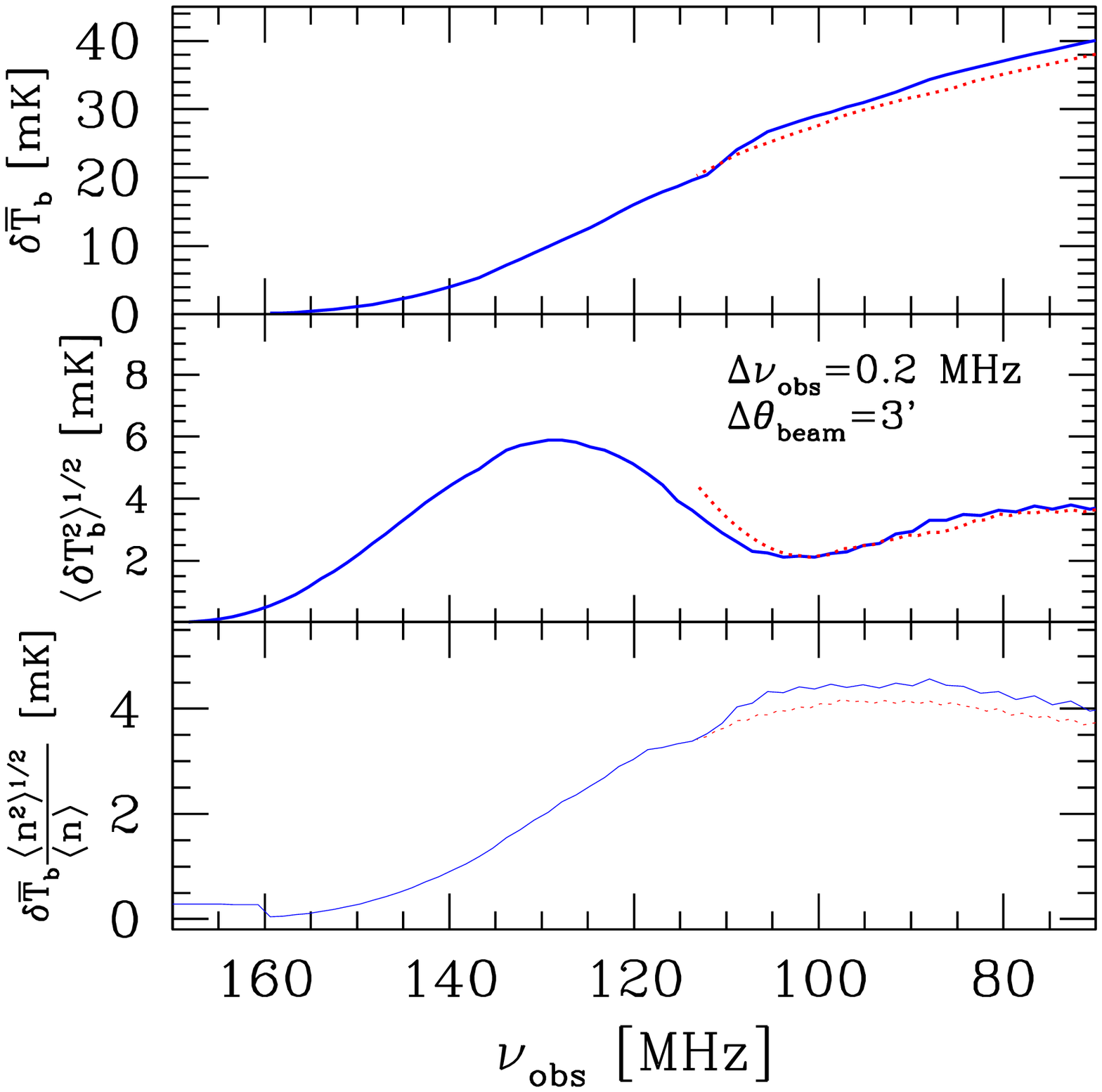}
  \includegraphics[height=1.89in]{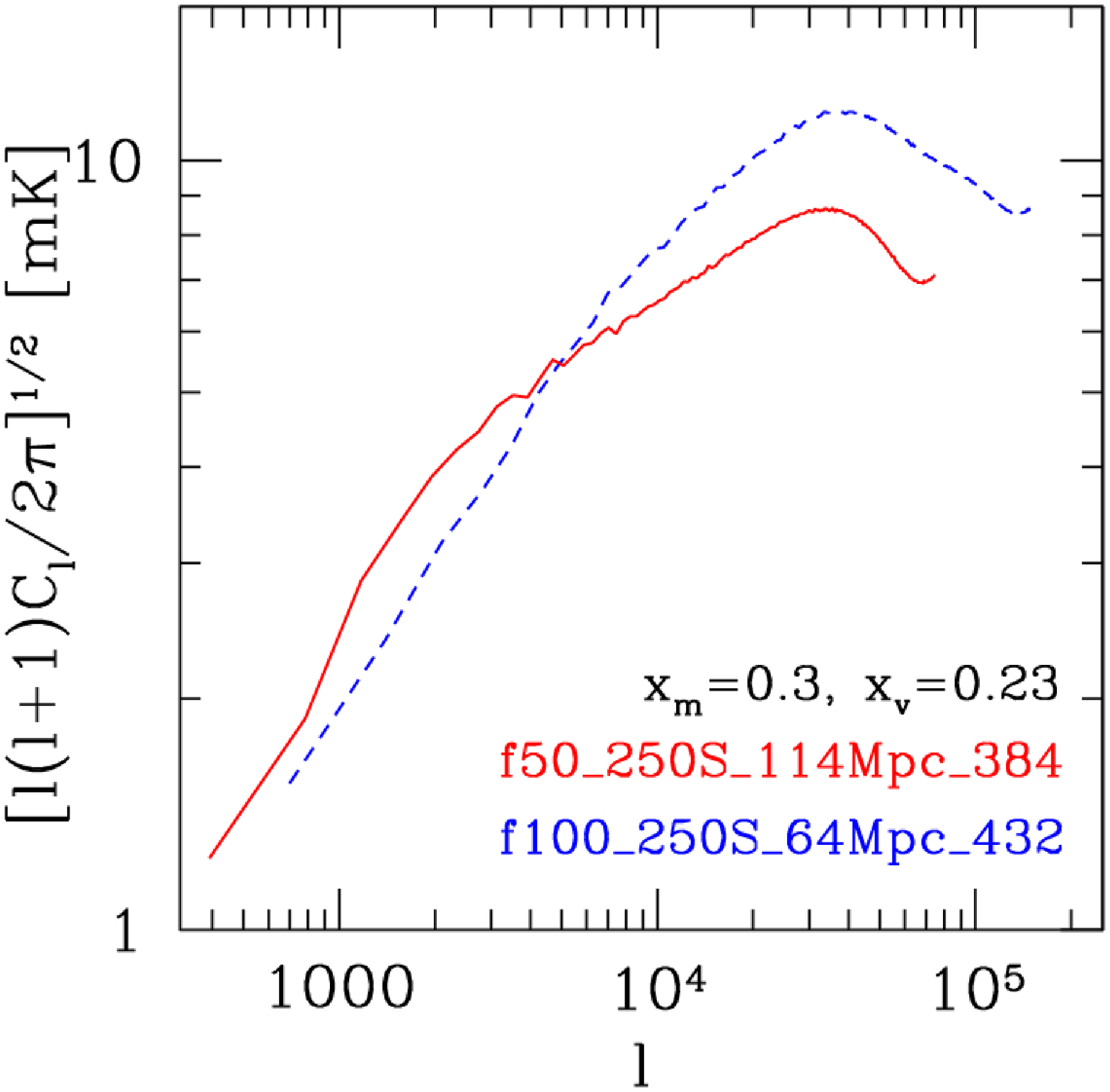}
  \caption{$21$-cm brightness temperature fluctuations (a) (left) rms
  fluctuations $\langle \delta T_b^2 \rangle^{1/2}$ [mK] versus mean
  ionized fraction $x_m$, for $91$ Mpc (solid) and $163$ Mpc (dotted)
  simulations, if observed with a beamwidth
  $\Delta \theta_\mathrm{beam} = 3'$ and bandwidth $\Delta
  \nu_\mathrm{obs} = 0.2$ MHz; (b) (middle) mean differential
  brightness temperature $\overline{\delta T_b}$ [mK] (top) and rms
  $\langle \delta T_b^2 \rangle^{1/2}$ [mK] (middle) versus observed
  frequency, and, for comparison, $\overline{\delta T_b} \langle n^2
  \rangle^{1/2}/\langle n \rangle$, to show that signal is enhanced
  relative to matter fluctuations, by reionization patchiness, and
  evolves differently; (c) (right) 2-D angular power spectrum of $\delta T_b$
  vs. $l$, for epoch at which $x_m=0.3$.} \label{threefigs}
\end{figure}

Predictions of the fluctuating 21-cm brightness temperature, $\delta
T_b$, from our 91 Mpc box ($\Delta\theta_{box} \sim 35'$ at $z \sim
8$) simulations are shown in Figure~\ref{skymap}, which maps the sky
along the line of sight in position - redshift (i.e. angle-frequency)
space.  The rms fluctuations shown in Figures~\ref{threefigs}(a) and
(b) for this case and the 163 box simulations, too, vary with
frequency as the ionization and density fluctuations evolve in
redshift.  There is a broad but distinct peak in the frequency range
$\Delta \nu \approx 120 -140$ Hz (i.e. x$_m \approx 0.5 -
0.8)$. Figure~\ref{threefigs}(c) also shows the 2-D angular power
spectrum of $\delta T_b$ when $x_m$ = 0.3, which peaks at $l \approx
3-4 \times 10^{4}$.  Qualitatively similar features have been seen
before from EOR simulations, but these are the first simulations which
are \textit{not only} of \textit{volumes large enough} to make
statistically reliable predictions, \textit{but also} of
\textit{mass-resolution high enough} to take explicit account of the
self-regulated nature of reionization.

% Fig. 5

\begin{theacknowledgments}
  This work was supported in part by NSF grant AST 0708176, NASA
  grants NNX07AH09G and NNG04G177G, Chandra grant SAO TM8-9009X, Swiss
  National Science Foundation grant 200021-116696/1, Swedish Research
  Council grant 60336701, and Texas Advanced Computing Center (TACC)
  computer allocation.
\end{theacknowledgments}

\bibliographystyle{aipprocl} % if natbib is missing

\bibliography{refs}

\begin{thebibliography}{10}
\providecommand{\enquote}[1]{``#1''}
\expandafter\ifx\csname url\endcsname\relax
  \def\url#1{\texttt{#1}}\fi
\expandafter\ifx\csname urlprefix\endcsname\relax\def\urlprefix{URL }\fi

\bibitem{2006PhR...433..181F}
S.~R. {Furlanetto}, S.~P. {Oh}, and F.~H. {Briggs}, \emph{\physrep}
  \textbf{433}, 181--301 (2006).

\bibitem{2006Sci...313..931B}
R.~{Barkana}, \emph{Science} \textbf{313}, 931--934 (2006).

\bibitem{2005SSRv..116..625C}
B.~{Ciardi}, and A.~{Ferrara}, \emph{Space Science Reviews} \textbf{116},
  625--705 (2005).

\bibitem{1994ApJ...427...25S}
P.~R. {Shapiro}, M.~L. {Giroux}, and A.~{Babul}, \emph{\apj} \textbf{427},
  25--50 (1994).

\bibitem{2006MNRAS.369.1625I}
I.~T. {Iliev}, G.~{Mellema}, U.-L. {Pen}, H.~{Merz}, P.~R. {Shapiro}, and M.~A.
  {Alvarez}, \emph{\mnras} \textbf{369}, 1625--1638 (2006).

\bibitem{2007ApJ...654...12Z}
O.~{Zahn}, A.~{Lidz}, M.~{McQuinn}, S.~{Dutta}, L.~{Hernquist},
  M.~{Zaldarriaga}, and S.~R. {Furlanetto}, \emph{\apj} \textbf{654}, 12--26
  (2007).

\bibitem{2007ApJ...657...15K}
K.~{Kohler}, N.~Y. {Gnedin}, and A.~J.~S. {Hamilton}, \emph{\apj} \textbf{657},
  15--29 (2007).

\bibitem{2007ApJ...671....1T}
H.~{Trac}, and R.~{Cen}, \emph{\apj} \textbf{671}, 1--13 (2007).

\bibitem{2006MNRAS.372..679M}
G.~{Mellema}, I.~T. {Iliev}, U.-L. {Pen}, and P.~R. {Shapiro}, \emph{\mnras}
  \textbf{372}, 679--692 (2006).

\bibitem{kSZ}
I.~T. {Iliev}, U.-L. {Pen}, J.~R. {Bond}, G.~{Mellema}, and P.~R. {Shapiro},
  \emph{\apj} \textbf{660}, 933--944 (2007).

\bibitem{2007arXiv0711.2944I}
I.~T. {Iliev}, P.~R. {Shapiro}, P.~{McDonald}, G.~{Mellema}, and U.-L. {Pen},
  \emph{ArXiv e-prints 0711.2994}  (2007).

\bibitem{2008MNRAS.384..863I}
I.~T. {Iliev}, G.~{Mellema}, U.-L. {Pen}, J.~R. {Bond}, and P.~R. {Shapiro},
  \emph{\mnras} \textbf{384}, 863--874 (2008).

\bibitem{2007MNRAS.376..534I}
I.~T. {Iliev}, G.~{Mellema}, P.~R. {Shapiro}, and U.-L. {Pen}, \emph{\mnras}
  \textbf{376}, 534--548 (2007).

\bibitem{teragrid08}
I.~T. {Iliev}, G.~{Mellema}, H.~{Merz}, P.~R. {Shapiro}, and U.-L. {Pen},
  \emph{TeraGrid08}  (2008), in press.

\bibitem{Hockney:1988:CSU}
R.~W. {Hockney}, and J.~W. {Eastwood}, \emph{{Computer simulation using
  particles}}, Bristol: Hilger, 1988.

\bibitem{2005NewA...10..393M}
H.~{Merz}, U.-L. {Pen}, and H.~{Trac}, \emph{New Astronomy} \textbf{10},
  393--407 (2005).

\bibitem{methodpaper}
G.~{Mellema}, I.~T. {Iliev}, M.~A. {Alvarez}, and P.~R. {Shapiro}, \emph{New
  Astronomy} \textbf{11}, 374 (2006).

\bibitem{comparison1}
I.~T. {Iliev}, B.~{Ciardi}, M.~A. {Alvarez}, A.~{Maselli}, A.~{Ferrara}, N.~Y.
  {Gnedin}, G.~{Mellema}, T.~{Nakamoto}, M.~L. {Norman}, A.~O. {Razoumov},
  E.-J. {Rijkhorst}, J.~{Ritzerveld}, P.~R. {Shapiro}, H.~{Susa}, M.~{Umemura},
  and D.~J. {Whalen}, \emph{\mnras} \textbf{371}, 1057--1086 (2006).

\bibitem{2006ApJ...647..397M}
G.~{Mellema}, S.~J. {Arthur}, W.~J. {Henney}, I.~T. {Iliev}, and P.~R.
  {Shapiro}, \emph{\apj} \textbf{647}, 397--403 (2006).

\bibitem{1974ApJ...187..425P}
W.~H. {Press}, and P.~{Schechter}, \emph{\apj} \textbf{187}, 425--438 (1974).

\bibitem{2002MNRAS.329...61S}
R.~K. {Sheth}, and G.~{Tormen}, \emph{\mnras} \textbf{329}, 61--75 (2002).

\end{thebibliography}

\end{document}